# Overview and Issues of Experimental Observation of Microbunching Instabilities

Alex H. Lumpkin
Mail to: lumpkin@fnal.gov
Fermi National Accelerator Laboratory, Batavia, IL 60510 USA

### 1.1.1 Introduction

The generation of the ultra-bright beams required by modern accelerators and drivers of free-electron lasers (FELs) has generally relied on chicane-based bunch compressions that often result in the microbunching instability [1,2]. Following compression, spectral enhancements extend even into the visible wavelengths through the longitudinal space charge (LSC) impedances. Optical transition radiation (OTR) screens have been extensively used for transverse electron beam size measurements for the bright beams, but the presence of such longitudinal microstructures (microbunching) in the electron beam or the leading edge spikes can result in strong, localized coherent enhancements (COTR) that mask the actual beam profile. Generally, we have observed effects in rf photocathode (PC) injected linacs with chicane compressions since an $R_{56}$ term is needed. In the past COTR had been only reported in S-band and L-band photoinjected based linacs with single or double bunch compression. Drive laser modulations and charge shot noise have been suspected of contributing to the cause. We now have evidence for the effects in both rf PC-gun injected linacs and thermionic-cathode (TC)-gun injected linacs (the latter do not involve a drive laser). Since the first observations, significant efforts have been made to characterize, model, and mitigate COTR effects on beam diagnostics [3-6]. An update on the state-of-the-art for diagnosing these effects will be given as illustrated by examples at the Linac Coherent Light Source (LCLS), Spring-8 Compact SASE Source (SCSS), Spring-8 Angstrom Compact Free Electron LAser (SACLA), Advanced Photon Source (APS), and the Next Linear Collider Test Accelerator (NLCTA). These observations continue to be of interest to the accelerator community.

### 1.1.2 Instability Effects

#### 1.1.2.1 Context

It should be kept in mind that the energy modulation amplitude is even stronger in the several-micron-period regime where it impacts the effective energy spread and can reduce FEL gain. As reference the original description by Saldin, Schneidmiller, and Yurkov [1] provides an analysis of the charge density noise being amplified via LSC impedances with the gain as a function of wavelength as shown in Fig. 1. In this case curve 1 includes energy spread as compared to curve 2 which is for a cold beam. Experimentally, one images the 0.4- to 0.7-µm regime of COTR with our standard CCD cameras in the various linac facilities. Another gain calculation has been given by Huang et al. [2] with maximum gain calculated at about 10 µm under an initial 150-µm period modulation with 8% amplitude as also shown in Fig. 1.

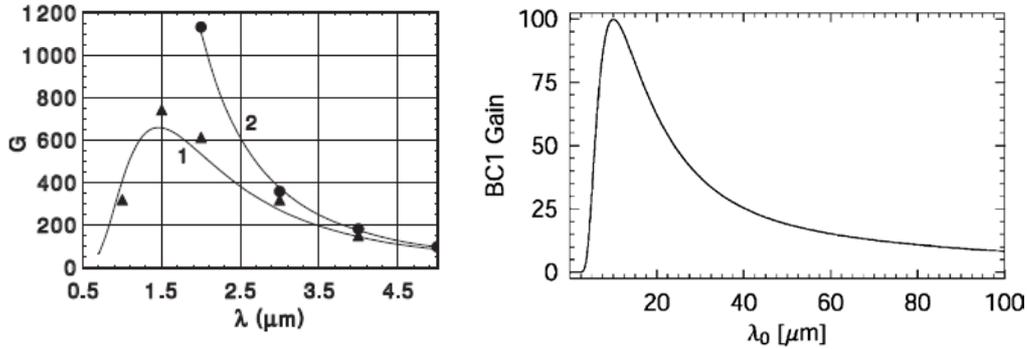

**Figure 1:** Calculated gain (G) for the microbunching instability versus wavelength from reference [1] (left) and reference [2] (right).

*1.1.2.2 Diagnostics Options*

Some of the diagnostics for assessing the µBI via COTR and other techniques include:
 1) Bunchlength monitors for tuning and verifying the compression. These might be based on coherent radiation sources based on transition radiation (CTR), synchrotron radaiation (CSR), edge radiation (CER), diffraction radiation (CDR), etc. in the frequency domain or on incoherent sources in the temporal domain with an ultrafast streak camera or deflecting mode cavity plus an imaging screen.
  2) OTR beam profile monitor screens are used for detecting the presence of COTR and its spatial distribution, intensity fluctuations, and intensity enhancements. The latter can be factors of 100 to 10,000 which make the profiles no longer representative of the true charge distribution and obviate the technique for profiling.
  3) Optical spectrometers have been used for characterizing the NIR COTR vs the bluish OTR. This information provides a concept for mitigation of the COTR effects in the diagnostics with spectral filtering.
  4) NIR and FIR spectrometers for evaluation of the spectral content in the 1-30 um regime. These experiments have predominately been done at FLASH [7,8].
  5) Deflecting mode cavities are needed with very high resolution (fs) to see directly the longitudinal structure even for FIR modulations. This is one diagnostics issue [6].
 6)  Electron energy spectrometers need high resolution to resolve the modulation [6].
 7) X-ray spectrometer with high resolution. There are direct spectral effects in the FEL spectra driven by such beams with modulations which can be detected as discussed below [9].
 These techniques have been applied on PC rf gun beams initially, but we have been able to apply them now to the TC gun beams as also will be described in a later section.
   The instability effects were graphically demonstrated in the high energy spectra at LCLS as presented at FEL 10 by J. Welch [9]. The modulation in energy attributed to such microbunching is seen with the laser heater off in Fig. 2a, while it is suppressed with the laser heater on in Fig. 2b. Concomitantly, the observed x-ray spectra for the two cases showed the dramatic simplification of the spectrum with laser heater "on" in Fig. 3.

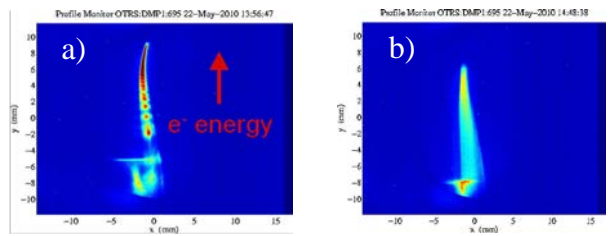

**Figure 2:** Examples of the LCLS electron beam high energy spectrum a) without and b) with the laser heater active [9].

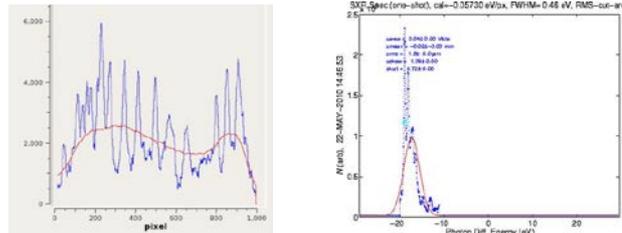

**Figure 3:** Corresponding x-ray spectra at LCLS for Fig.2 without (left) and with (right) the laser heater active [9].

### 1.1.3 Thermionic Cathode Gun Beams

One of the major developments in the past year in this µBI subfield involves the observations of the COTR effects attributed to the microbunching instability in TC gun beams, in both DC and rf guns. Example results are provided in this section.

#### 1.1.3.1 SCSS Results

The SCSS linac is based on a DC TC gun with deflector, subharmonic bunchers, a S-band accelerator section, a chicane bunch compressor, a C-band accelerator and another chicane for filtering the dark current beam as shown in Fig. 4 [10]. Using the second chicane as a bunch compressor was suggested in discussions at the µBI-4 workshop and following the FEL12 conference as well as looking for OTR enhancements at the station after this chicane. The experiments were initiated in October 2012 and were immediately successful.

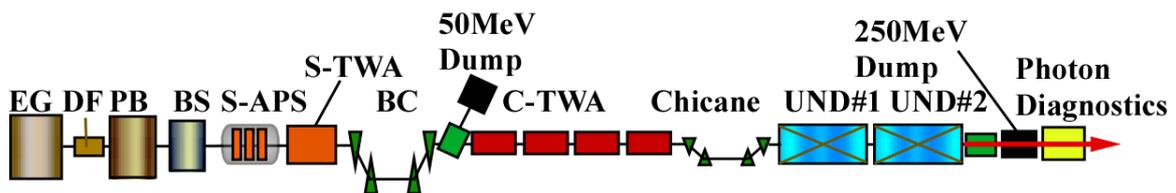

**Figure 4:** Schematic of the injector for the SCSS facility showing TC DC gun, accelerators, FEL, and beamlines. (courtesy of K. Togawa.)

An example image is shown in Fig. 5 with about 250 pC micropulse charge, and the plot of OTR intensity in such images versus C-band phase setting is shown in Fig. 6. The intensity doubles at the -10 degrees off-crest phase point, and the fluctuations of the intensity dramatically increase compared to those at -5 and -15 degrees. This increase is attributed to a coherent process starting from noise in the beam due to the LSC microbunching instability, and the second compression shifted the effects into the visible light regime where they were sensed by the CCD camera. In the previous tests they had observed the OTR *before* this second chicane and with the C-band accelerator run on crest so no COTR effects were observed.

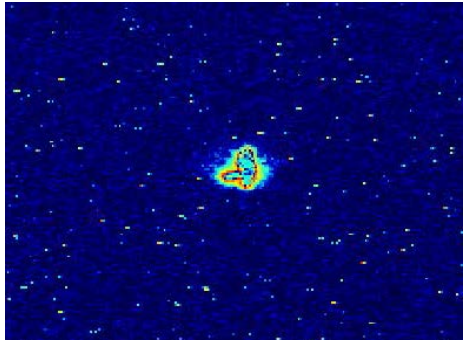

**Figure 5:** Image of the OTR and COTR generated at the beam profile station after the second bunch compressor at SCSS [11]**.** The x and y axes cover 4 x 3 mm.

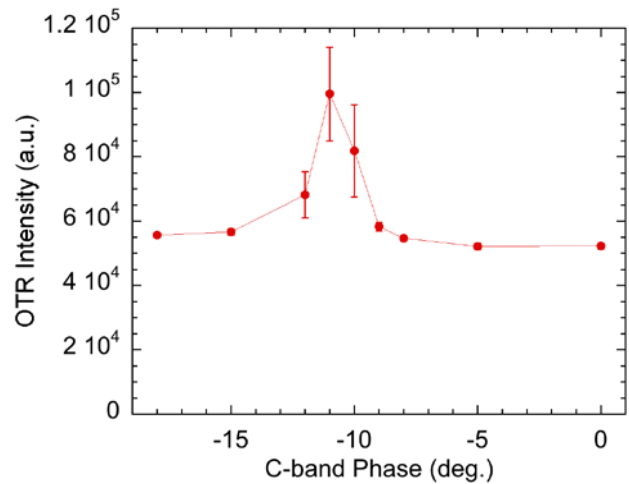

**Figure 6:** Plot of the C-band phase dependence of the OTR from the station after the second chicane at SCSS [11].

*1.1.3.2 SACLA Results*

The SACLA accelerator injector shown in Fig. 7 is based on the SCSS design with a few modifications such as the C-band correcting cavity. It also uses a DC TC gun with a deflector to select a part of the beam that is then subharmonically bunched. There are then three chicane-based compression stages with further acceleration to 1.4 GeV. After the third chicane bunch compression, they encountered significant COTR enhancements that saturated their CCD cameras [12]. To mitigate this effect they used spatial filtering with a scintillator crystal to obtain beam images. However, in the last year the staff revisited the stations to quantify the effects per suggestions from the Microbunching Instability Workshop 2012 attendees.

SACLA staff now report that the enhancement, or gain, is about 6000 over OTR [11], and they also showed the characteristic gradient-operator-related doughnut shape in the near field beam image in Fig. 8 as described by Loos et al. previously in the LCLS COTR images [4]. Additionally, they reported the enhanced red wavelength regime with intensity modulated spectrum in Fig. 9 as identified in the earlier APS/ANL PC rf gun based linac studies [13].

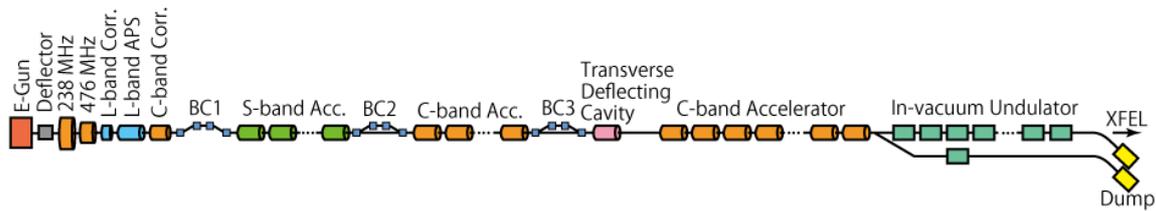

**Figure 7:** Schematic of the SACLA beamline with DC gun, bunchers, and accelerators with three chicanes for bunch compresson. (courtesy of K. Togawa).

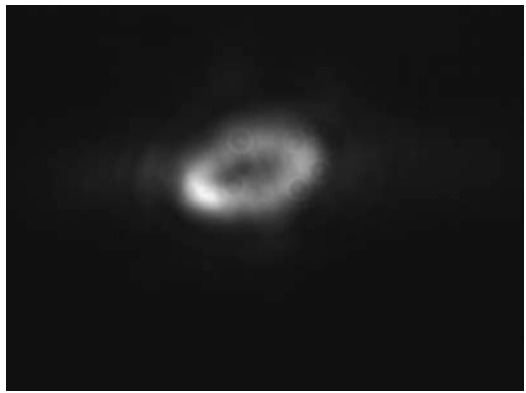

**Figure 8:** Beam image at 1.4 GeV after the third chicane at SACLA showing the COTR halo in the near field [11].

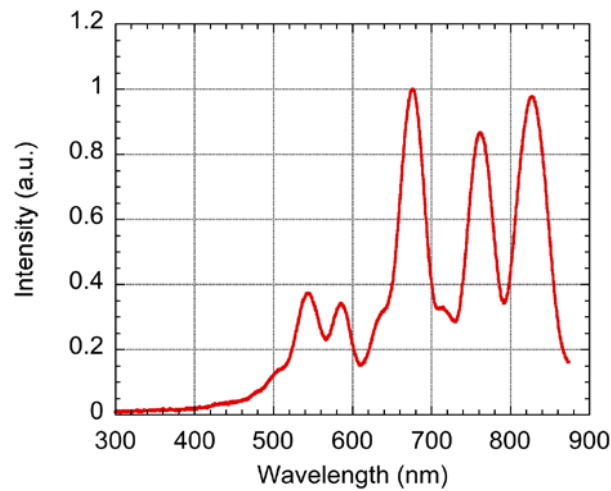

**Figure 9:** COTR spectrum obtained at 1.4 GeV after the third chicane at SACLA [11].

*1.1.3.3 APS/ANL Results*

The initial experiments at ANL were on the PC rf gun beam, and the first look at the TC rf gun beam was also done. At that time the signature of COTR spiking in the beam profiles was only seen in the PC rf gun OTR images [5]. However, because the TC rf gun beam involves a set of 25 micropulses at the S-band frequency, the statistical fluctuations of COTR might be averaged out in the CCD camera integration. More recent tests show the increase in the integrated profiles when the compression in the chicane occurs following implementation of energy chirp in the beam entering the chicane. In these cases we operated with higher current in the gun than previously. Adjusting the compression was done by evaluating the autocorrelations of FIR CTR at stations located in the linac before and after the chicane as shown in Fig. 10. The alpha magnet did provide an initial compression of about ten prior to the chicane's factor of two compression measured. As shown in Fig. 11, two different horizontal profiles from 10 OTR-image sums were taken without (black and green circles) and with chicane compression (blue and red circles). The intensity of the profile peaks increased by 4-8 when operating at the rf phase that peaked the FIR CTR signal in the Golay cell after the chicane [7]. The profile data shown at the workshop have been replotted in this figure to facilitate the direct comparison of the intensities. The charge transport at the end of the linac was tracked at 2 nC ±10% during the acquisition of these sets of images so charge-transport variation cannot explain the effects.

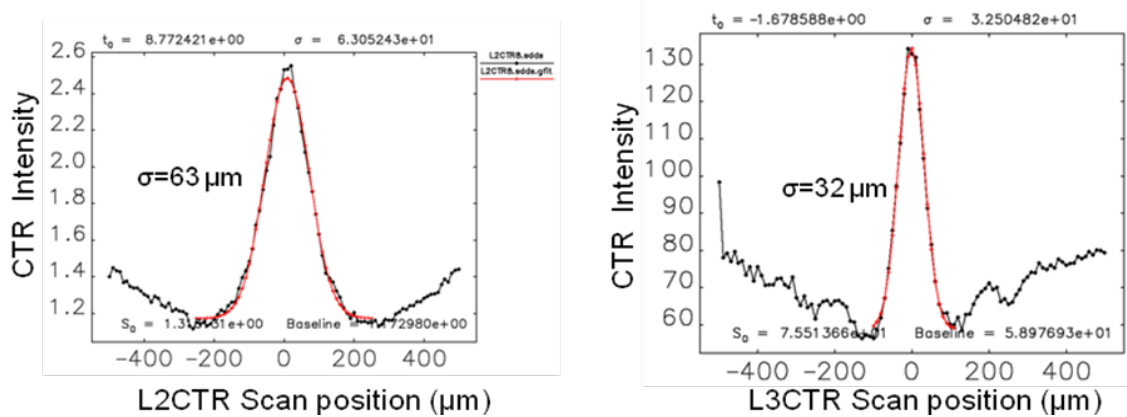

**Figure 10:** Autocorrelation results of CTR taken at the L2CTR and L3CTR locations which are before and after the chicane, respectively. A compression factor of two was observed.

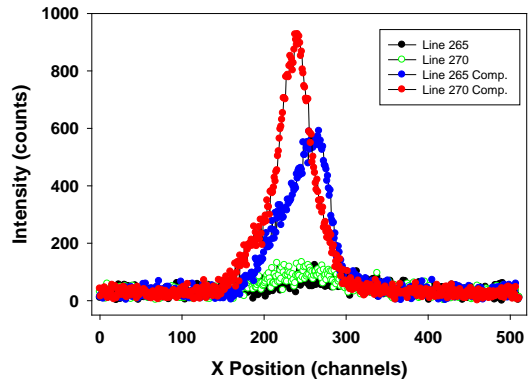

**Figure 11:** Profiles at lines 265 and 270 through the OTR sum images from TC rf gun beam uncompressed and compressed (blue and red symbols) with a final beam energy of 325 MeV at APS.

### 1.1.4  *NLCTA X-band RESULTS*

Another interesting piece of the puzzle involves the observation of COTR with only 20-pC of charge in the micropulse following two chicane compressions at NLCTA as shown in Fig. 12. This facility has an S-band PC rf gun with two X-band accelerator sections that produce the 120-MeV beams [14]. Additionally, coherent optical undulator radiation has been reported [15].

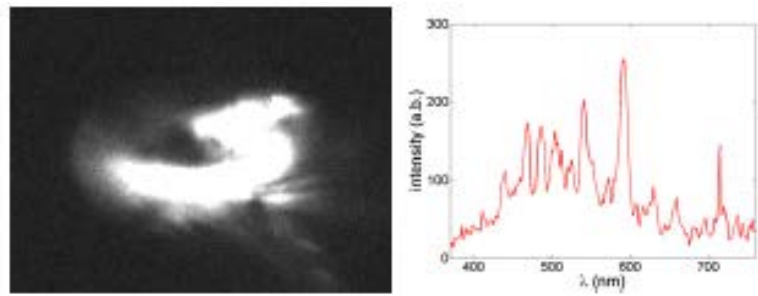

**Figure 12:** Typical COTR image (left) and wavelength spectrum following double compression at NLCTA [14].

### 1.1.5 Discussion

Table 1 is a summary of the scope of the observations in the various linacs including LCLS, DESY, and NLCTA and with the new TC gun beam results at SCSS, SACLA, and APS. The role of compression factors is indicated where second compressions in SCSS and APS were needed to display the COTR effect in TC gun beams. It is noted that the final enhancement of 6000 in SACLA after three chicanes in the TC DC gun beam approaches the very large enhancements in LCLS after two chicane compressions of the PC rf gun beam. Also, it is noted the transverse normalized emittances vary from 6-10 mm mrad in APS and NLCTA beams while LCLS, DESY, SCSS, and SACLA beams have emittances at about 1 mm mrad or below. All cases below exhibit some COTR effects.

Table 1: Summary of the COTR Effects Observed in Various Accelerator Facilities Including the Gun Type, Linac Energy, and Number of Chicanes or Compressions.

| Facility | Gun | Linac, Energy | Chicanes | COTR Effects |
|---|---|---|---|---|
| LCLS | PC, S-band | S-band, 250, 14 GeV | two | very strong, x$10^4$ |
| APS | PC, S-band rf TC, S-band | S-band, 150, 325 MeV | one alpha magnet, one | x10-100 localized x4 integral |
| DESY | PC, L-band | SCRF, L-band, 1.2 GeV, linearizer | two | x 10-100 localized |
| SACLA | TC, DC gated | S-band, C-band, 1.4 GeV | three | >6x$10^3$ after 3 compressions |
| SCSS | TC, DC gated | S-band , C-band, 250 MeV | two | x2, Observable after two compressions |
| NLCTA | PC, S-band | X-band, 120 MeV | two of four | x20 after two |

**1.1.6 Some Issues**

There are some issues on the experimental side to consider in the future. These include:
  -extending measurements in the NIR and FIR where the instability gain is stronger in more of the accelerator configurations;
  -obtaining detailed longitudinal measurements with adequate resolution in time and energy;
  - defining of beam parameters needed for simulations;
  -evaluating longitudinal impedances involved in C-band and X-band accelerating structures compared to those of L-band and S-band structures;
  - collecting more statistical data on intensity fluctuations and the Gamma function for the process;
  - benchmarking of the relevant codes with the more extensive data sets we now have.

The question of whether the PC gun beams have more charge fluctuations than TC gun beams, and hence they are more prone to the larger μBI effects needs consideration.

**1.1.7 Summary**

In summary, the microbunching instability as detected through the generation of COTR has become worldwide in interest. The observations of the microbunching instability attributed to longitudinal space charge impedances and CSR effects has become recognized as a more general phenomenon with cases reported in L-band, S-band, C-band, and X-band accelerators and with beams generated by both PC rf guns and TC rf and DC guns. There is an opportunity for using this broader empirical data base to elucidate the effect via further modeling efforts based on LSC impedances and shot noise. Modeling of the TC gun beams still seems to be needed since the slice energy spread may not be as well understood at this time. Mitigations in the diagnostics have been reported in several labs [5,6,12], and suppression of the instability itself has been ongoing with laser heaters and dispersive elements. Further investigations are encouraged as it has now been demonstrated that the instability is not only observed in PC rf gun beams initiated with drive lasers as was implied a few workshops ago.

**1.1.8 Acknowledgments**

The author acknowledges: discussions with D. Dowell, Z. Huang, and H. Loos of SLAC; W. Fawley (LBNL); D. Rule (NSWC); and R. Fiorito (UMD); support from K.-J. Kim, R. Gerig, and H. Weerts of the Argonne Accelerator Institute; the collaborations on ANL linac tests of N. Sereno, J. Dooling, S. Pasky, Y. Li, and Y. Sun of ANL; the support of M.Wendt and N. Eddy of Fermilab; and the figures provided by K. Togawa (Riken,SPring-8) and S. Weathersby (NLCTA/SLAC).